\documentstyle[aps,preprint,amsmath,eucal,amsthm,amssymb,latexsym]{revtex}

\DeclareFontFamily{U}{msb}{}
\DeclareFontShape{U}{msb}{m}{n}{
<5><6><7><8><9> gen *msbm <10><10.95><12><14.4><17.28><20.74><24.88>msbm10}{}
\DeclareSymbolFont{AMSb}{U}{msb}{m}{n}
\DeclareMathSymbol{\bA}{\mathbin}{AMSb}{'101}
\DeclareMathSymbol{\bB}{\mathbin}{AMSb}{'102}
\DeclareMathSymbol{\bC}{\mathbin}{AMSb}{'103}
\DeclareMathSymbol{\bD}{\mathbin}{AMSb}{'104}
\DeclareMathSymbol{\bE}{\mathbin}{AMSb}{'105}
\DeclareMathSymbol{\bF}{\mathbin}{AMSb}{'106}
\DeclareMathSymbol{\bG}{\mathbin}{AMSb}{'107}
\DeclareMathSymbol{\bH}{\mathbin}{AMSb}{'110}
\DeclareMathSymbol{\bI}{\mathbin}{AMSb}{'111}
\DeclareMathSymbol{\bJ}{\mathbin}{AMSb}{'112}
\DeclareMathSymbol{\bK}{\mathbin}{AMSb}{'113}
\DeclareMathSymbol{\bL}{\mathbin}{AMSb}{'114}
\DeclareMathSymbol{\bM}{\mathbin}{AMSb}{'115}
\DeclareMathSymbol{\bN}{\mathbin}{AMSb}{'116}
\DeclareMathSymbol{\bO}{\mathbin}{AMSb}{'117}
\DeclareMathSymbol{\bP}{\mathbin}{AMSb}{'120}
\DeclareMathSymbol{\bQ}{\mathbin}{AMSb}{'121}
\DeclareMathSymbol{\bR}{\mathbin}{AMSb}{'122}
\DeclareMathSymbol{\bS}{\mathbin}{AMSb}{'123}
\DeclareMathSymbol{\bT}{\mathbin}{AMSb}{'124}
\DeclareMathSymbol{\bU}{\mathbin}{AMSb}{'125}
\DeclareMathSymbol{\bV}{\mathbin}{AMSb}{'126}
\DeclareMathSymbol{\bW}{\mathbin}{AMSb}{'127}
\DeclareMathSymbol{\bX}{\mathbin}{AMSb}{'130}
\DeclareMathSymbol{\bY}{\mathbin}{AMSb}{'121}
\DeclareMathSymbol{\bZ}{\mathbin}{AMSb}{'132}

\newtheoremstyle{definition}%
{6pt}{6pt}{\itshape}%
{}{\bfseries}{.}{ }{}
\theoremstyle{definition}

\def\RLD#1#2{\mbox{\rm D}^{#1}_{#2}}

\def\RLI#1#2{\mbox{\rm I}^{#1}_{#2}}

\def\LT#1{{\cal L}\left\{#1\right\}}
\def\FT#1{{\mathcal F}\left\{#1\right\}}

\def\MT#1{{\cal M}\left\{#1\right\}}

\def\vek#1{\pmb{#1}}
\def\HH#1#2{\left|\begin{array}{l}{#1}\\[1ex]{#2}\end{array}\right.}

\def\d{\mbox{\rm d}}

\DeclareMathOperator{\Rea}{Re}

\def\L1loc{L^1_{\rm loc}}

\def\al{\alpha}

\def\ga{\gamma}

\def\la{\lambda}
\def\ta{\tau}

\def\Ga{\Gamma}

\def\toi{\to\infty}

\def\fo{\al}

\def\Lc{{\cal L}}

\def\dst{\displaystyle}

\def\DC{C}

\begin{document}
\draft

\setcounter{page}{0}
\title{Fractional Diffusion based on\\Riemann-Liouville Fractional Derivatives}
\author{R. Hilfe$\mbox{\rm r}^{1,2}$}
\address{
$\mbox{ }^1$ICA-1, Universit{\"a}t Stuttgart,
Pfaffenwaldring 27, 70569 Stuttgart\\
$\mbox{ }^2$Institut f{\"u}r Physik,
Universit{\"a}t Mainz,
55099 Mainz, Germany}
\maketitle
\thispagestyle{empty}
\vspace*{2cm}
\begin{center}
{\em Dedicated to Harvey Scher on the occasion of his 60th birthday.}
\end{center}
\begin{abstract}
A fractional diffusion equation based on Riemann-Liouville
fractional derivatives is solved exactly.
The initial values are given as fractional integrals.
The solution is obtained in terms of $H$-functions.
It differs from the known solution of fractional
diffusion equations based on fractional integrals.
The solution of fractional diffusion based on
a Riemann-Liouville fractional time derivative
does not admit a probabilistic interpretation
in contrast with fractional diffusion based on
fractional integrals.
While the fractional initial value problem is well
defined and the solution finite at all times
its values for $t\to 0$ are divergent.
\end{abstract}
\vspace*{1cm}
{\tt J.Phys.Chem B, vol. 104, page 3914, (2000)}
\newpage

Anomalous subdiffusive transport appears to be a universal
experimental phenomenon \cite{BKZ86,shl88,wei88}.
Examples occur in widely different systems ranging
from amorphous semiconductors \cite{SM75,ERJMR84} through
polymers \cite{cat84b,sta84,SB95} and composite heterogeneous
films \cite{NG86} to porous media \cite{BS97,KMK97}.
Theoretical investigations into anomalous
diffusion and continuous time random walks 
have been a major focus of H. Schers research
for many years \cite{MS73,SL73,SM75,BS97}.
The purpose of this paper is to discuss
a theoretical approach based on the replacement
of the time derivative in the diffusion equation 
with a derivative of noninteger order (fractional derivative).

Many investigators have proposed 
the use of fractional time derivatives for subdiffusive
transport on a purely mathematical or heuristic basis 
\cite{BW68,nig86,SW89,hil89c,non90,jum92,SB95,hil95e}.
From the perspective of theoretical physics
this proposal touches upon fundamental
principles such as locality, irreversibility and
invariance under time translations because fractional 
derivatives are nonlocal operators that are not invariant 
under time reversal \cite{hil98e}. 
These issues are generally avoided in heuristic
and mathematical proposals, but were discussed 
recently in the context of long time limits and
coarse graining \cite{hil98e}.
It was found that fractional 
time derivatives with orders between $0$ and
$1$ may generally appear as infinitesimal 
generators of a coarse grained macroscopic time 
evolution \cite{hil95e,hil98e,hil93e,hil95c,hil95f}.

Differential equations involving fractional
derivatives raise a second basic problem, related
to the first, that will be the focus of this paper.
The second problem is whether to replace the integer
order derivative by a Riemann-Liouville, by a Weyl,
by a Riesz, by a Gr{\"u}nwald or by a Marchaud fractional
derivative (see \cite{BT68,ros75,hil97b,BW00} for definitions
of these different derivatives).
Different authors have introduced different
derivatives depending on the physical situation
\cite{hil98e,NM99,SFB00,WG99}.

Given the basic objective of introducing
fractional derivatives into the diffusion
equation the present paper will be concerned with
the equation
\begin{equation}
\RLD{\al}{0+}f(\vek{r},t)=\DC_\al\Delta f(\vek{r},t)
\label{fdd}
\end{equation}
where $f(\vek{r},t)$ denotes the unknown field
and $\DC_\al$ denotes the fractional diffusion
constant with dimensions $[{\rm cm/s}^{\al}]$.
The fractional derivative operator, denoted
as $\RLD{\al}{0+}$, is the Riemann-Liouville 
derivative of order $\al$ and with lower limit $a\in\bR$.
It is defined as \cite{BT68,BW00}
\begin{equation}
(\RLD{\al}{a+}f)(x)=\frac{\d}{\d x}(\RLI{1-\al}{a+}f)(x)
\end{equation}
where 
\begin{equation}
(\RLI{\al}{a+}f)(x) = \frac{1}{\Ga(\al)}
\int_a^x (x-y)^{\al-1}f(y)\;\d y
\label{RLI}
\end{equation}
is the  Riemann-Liouville fractional integral with order 
$\al$ and lower limit $a$.
Although many authors have investigated
fractional diffusion problems 
\cite{BW68,nig86,wys86,SW89,jum92,hil95b}
it seems that equation (\ref{fdd}) has not been 
solved previously.
In fact it was recently questioned
whether an approach using eq. (\ref{fdd}) is consistent
\cite{MBK99,NM99}.
It is the purpose of this paper to solve eq. (\ref{fdd})
exactly thereby establishing its consistency for
appropriate initial conditions.

Let me emphasize that eq. (\ref{fdd}) differs from 
the popular equation introduced and solved in \cite{SW89}.
The latter equation is obtained by 
first rewriting the diffusion equation in integral form as
\begin{equation}
f(\vek{r},t)=f_0\delta(\vek{r})+\DC_1\int_0^t\Delta f(\vek{r},t')\;\d t'
\label{diffeq}
\end{equation}
where $\DC_1$ is the usual diffusion constant,
$\delta(\vek{r})$ is the Dirac measure at the origin, 
and where the initial condition $f(\vek{r},0)=f_0\delta(\vek{r})$
has been incorporated.
Then the integral on the right hand side is
replaced by a fractional Riemann-Liouville
integral to arrive at the fractional integral form
\begin{subequations}
\label{fdi}
\begin{equation}
f(\vek{r},t)=f_0\delta(\vek{r})+\frac{\DC_\al}{\Ga(\al)}
\int_0^t(t-t')^{\al-1}\Delta f(\vek{r},t')\;\d t'
=f_0\delta(\vek{r})+\DC_\al(\RLI{\al}{0+}\Delta f)(\vek{r},t)
\label{fdia}
\end{equation}
or, upon differentiating both sides, at 
\begin{equation}
\frac{\partial}{\partial t}f(\vek{r},t)=
\DC_\al(\RLD{1-\al}{0+}\Delta f)(\vek{r},t)
\label{fdib}
\end{equation}
\end{subequations}
where $\DC_\al$ is again a fractional diffusion constant.
For $\al=1$ this reduces to eq. (\ref{diffeq}).
The exact solution of eq. (\ref{fdi}) is known
and given by eq. (\ref{usual}) below.

In \cite{hil95a,hil98c} it was shown that eqs. (\ref{fdi})
have a rigorous relation with continuous time random walks
of the kind investigated frequently by Harvey Scher
\cite{MS73,SL73,SM75,BS97}.
More precisely, eq. (\ref{fdd}) was found to
correspond exactly to a continuous time random
walk with the long tailed waiting time density
\begin{equation}
\psi(t;\al,\ta_0)=
\frac{1}{\ta_0}\left(\frac{t}{\ta_0}\right)^{\al-1}
E_{\al,\al}\left(-\frac{t^\al}{\ta_0^\al}\right)
\end{equation}
where $\ta_0$ is a time constant.
Here $E_{a,b}(x)$ denotes the generalized Mittag-Leffler function
defined by
\begin{equation}
E_{a,b}(x) = \sum_{k=0}^\infty \frac{x^k}{\Ga(a k +b)}
\label{Mittag-Leffler}
\end{equation}
for all $a>0$ and $b\in\bC$.
For $\al=1$ this reduces to an exponential waiting time density.
For $0<\al<1$ these waiting time densities have a long tail
decaying as $\psi(t)\sim t^{-1-\fo}$ for $t\toi$.
Interestingly $\psi(t)\sim t^{\fo -1}$ diverges algebraically
for $t\to 0$.
It follows from refs. \cite{hil93e,hil95c,hil95e,hil95f,hil98e}
that among the the waiting time densities with long tails the
densities $\psi(t;\al,\ta_0)$ represent important universality 
classes for continuous time random walks.

Note that eqs. (\ref{fdi}) and (\ref{fdd})
are not equivalent.
The difference between eqs. (\ref{fdi}) and (\ref{fdd})
has to do with the initial conditions.
An appropriate inital condition is found
by analysing the stationary case.
One finds that the fractional integral
\begin{equation}
\RLI{(1-\fo)}{0+}f(\vek{r},0+)=f_{0,\al}\delta(\vek{r})
\label{initial}
\end{equation}
is preserved during the time evolution.
This is a nonlocal initial condition.
It implies the divergence of $f(\vek{r},t)$
as $t\to 0$, as is characterstic for
fractional stationarity \cite{hil95e,hil95f}.

Equation (\ref{fdd}) with initial condition (\ref{initial})
can be solved exactly by Fourier-Laplace techniques.
Let the Fourier transformation be defined as
\begin{equation}
\FT{f(\vek{r})}(\vek{q})=
\int_{\bR^d} e^{i\vek{q}\cdot\vek{r}}f(\vek{r})\d\vek{r} .
\end{equation}
Fourier and Laplace transformation of eq. (\ref{fdd}) now yield
\begin{equation}
f(\vek{q},u) = \frac{f_{0,\al}}{\DC_\al\vek{q}^2+u^\fo}.
\label{diflapfou}
\end{equation}
Inverting the Laplace transform gives 
\begin{equation}
f(\vek{q},t) = f_{0,\al}\;t^{(\fo-1)}E_{\fo,\fo}
(-\DC_\al\vek{q}^2t^\fo).
\label{diffou}
\end{equation}
Setting $\vek{q}=0$ shows that $f(\vek{r},t)$ cannot be a probability
density because its normalization would depend on $t$.
Hence eq. (\ref{fdd}) does not admit a probabilistic interpretation
contrary to eq. (\ref{fdi}).

To obtain $f(\vek{r},t)$ it is advantageous to first invert 
the Fourier transform in eq. (\ref{diflapfou}) and only later
the Laplace transform.
The Fourier transform may be inverted by noting the formula
\cite{BN71}
\begin{equation}
(2\pi)^{-d/2}\int e^{i\vek{q}\cdot\vek{r}}
\left(\frac{|\vek{r}|}{m}\right)^{1-(d/2)}
K_{(d-2)/2}\left(m|\vek{r}|\right)\;\d\vek{r}
=\frac{1}{\vek{q}^2+m^2}
\end{equation}
which leads to
\begin{equation}
f(\vek{r},u)= f_{0,\al}(2\pi \DC_\al)^{-d/2}\left(\frac{r}{\sqrt{\DC_\al}}\right)^{1-(d/2)}
u^{\fo(d-2)/4}
K_{(d-2)/2}\left(\frac{ru^{\fo/2}}{\sqrt{\DC_\al}}\right)
\end{equation}
with $r=|\vek{r}|$.
To invert the Laplace transform it is convenient to use
the relation
\begin{equation}
\MT{f(t)}(s) = \frac{\MT{\LT{f(t)}(u)}(1-s)}{\Ga(1-s)}
\label{laplmell}
\end{equation}
between the Laplace transform and the Mellin transform
\begin{equation}
\MT{f(t)}(s)=\int_0^\infty t^{s-1}f(t)\;\d t .
\label{mellin}
\end{equation}
of a function $f(t)$.
Setting $A=r/\sqrt{\DC_\al}$, $\la=\fo/2$, $\nu=(d-2)/2$ and
$\mu=\fo(d-2)/4$ and using the general
relation
\begin{equation}
\MT{x^qg(bx^p)}(s)=\frac{1}{p}b^{-(s+q)/p}g\left(\frac{s+q}{p}\right)
\qquad (b,p>0)
\end{equation}
leads to
\begin{equation}
\MT{f(r,u)}(s)=\frac{f_{0,\al}}{\la}(2\pi \DC_\al)^{-d/2}
A^{1-(d/2)}A^{-(s+\mu)/\la}
\MT{K_\nu(u)}\left((s+\mu)/\la\right).
\end{equation}
The Mellin transform of the Bessel function reads \cite{obe74}
\begin{equation}
\MT{K_\nu(x)}(s) = 2^{s-2}\Ga\left(\frac{s+\nu}{2}\right)
\Ga\left(\frac{s-\nu}{2}\right).
\end{equation}
Inserting this, using eq.(\ref{laplmell}), and restoring the
original variables then yields
\begin{equation}
\MT{f(r,t)}(s) =
\frac{f_{0,\al}}{\fo(r^2\pi)^{d/2}}
\left(\frac{r}{2\sqrt{\DC_\al}}\right)^{2(1-(1/\fo))}
\left(\frac{r}{2\sqrt{\DC_\al}}\right)^{2s/\fo}
\frac{\Ga\left(\frac{d}{2}-(1-\frac{1}{\fo})-\frac{s}{\fo}\right)
\Ga\left(\frac{1}{\fo}-\frac{s}{\fo}\right)}
{\Ga(1-s)}
\end{equation}
for the Mellin transform of $f$.
Comparing this with the definition of the general $H$-function
in eqs. (\ref{Hdef}),(\ref{Hkrn}) allows to identify the 
$H$-function parameters as $m=0$,$n=2$,$p=2$,$q=1$, $A_1=A_2=1/\fo$,
$a_1=1-(d/2)+(1-(1/\fo))$, $a_2=(1-(1/\fo))$,
$b_1=0$ and $B_1=1$ if $(\fo d/2)-(\fo-1)>0$.
Then the result becomes
\begin{align}
&f(r,t) = \frac{f_{0,\al}}{\fo(r^2\pi)^{d/2}}
\left(\frac{r}{2\sqrt{\DC_\al}}\right)^{2(1-(1/\fo))}\nonumber\\
&H^{02}_{21}\left(\left(\frac{2\sqrt{\DC_\al}}{r}\right)^{2/\fo}t
\HH{(1-\frac{d}{2}+(1-\frac{1}{\fo}),\frac{1}{\fo}),
((1-\frac{1}{\fo}),\frac{1}{\fo})}{(0,1)}
\right) .
\end{align}
This may be simplified using eqs.(\ref{Hreciprocal}), 
(\ref{Hscalarmult}) and (\ref{Hpowermult}) to become finally
\begin{equation}
f(r,t) = \frac{f_{0,\al}\;t^{(\fo-1)}}{(r^2\pi)^{d/2}}
H^{20}_{12}\left(\frac{r^2}{4\DC_\al t^\fo}
\HH{(\fo,\fo)}{(d/2,1),(1,1)}
\right) .
\end{equation}
This result should be compared with the known solution
\begin{equation}
f(r,t) = \frac{f_0}{(r^2\pi)^{d/2}}
H^{20}_{12}\left(\frac{r^2}{4\DC_\al t^\fo}
\HH{(1,\fo)}{(d/2,1),(1,1)}
\right)
\label{usual}
\end{equation}
of eq. (\ref{fdi}) in which case $f(\vek{r},t)$
is also a probability density.

In summary this paper has shown that fractional diffusion
based on Riemann-Liouville derivatives requires a fractional
inital condition given by eq. (\ref{initial}).
With this initial condition the fractional Cauchy problem
can be solved exactly in terms of $H$-functions, and the
solution is similar to the exact solution of the fractional 
integral form in eq. (\ref{fdi}).
However contrary to eq. (\ref{fdi}) whose solution is a 
probability density, and which is related to continuous 
time random walks, the solution of eq. (\ref{fdd})
does not admit a probabilistic interpretation.

The reader may ask why it is important for theoretical
physics and chemistry to investigate different forms of
fractional diffusion.
An answer was given already in
\cite{hil95e,hil98e,hil93e,hil95c,hil95f},
and recently again in \cite{hil00a,hil00b}.
In these works it was found that fractional 
time derivatives arise generally as infinitesimal 
generators of the time evolution when taking
a long-time scaling limit.
Hence the importance of investigating fractional
equations arises from the necessity to sharpen the 
concepts of equilibrium, stationary states and time 
evolution in the long time limit.

\section*{Appendix: H-Functions}
The $H$-function of order $(m,n,p,q)\in\bN^4$ and parameters
$A_i\in\bR_+ (i=1,\ldots,p)$, $B_i\in\bR_+ (i=1,\ldots,q)$, 
$a_i\in\bC (i=1,\ldots,p)$, and $b_i\in\bC (i=1,\ldots,q)$
is defined for $z\in\bC,z\neq 0$ by the contour integral
\cite{fox61,bra64,MS78,SGG82,PBM90}
\begin{equation}
H^{m,n}_{p,q}\left(z
\HH{(a_1,A_1),\ldots,(a_p,A_p)}{(b_1,B_1),\ldots,(b_q,B_q)}
\right) =
\frac{1}{2\pi i}\int_\Lc \eta(s)
z^{-s}\;\d s
\label{Hdef}
\end{equation}
where the integrand is
\begin{equation}
\eta(s)=
\frac{\dst\prod_{i=1}^m\Ga(b_i+B_is)\prod_{i=1}^n\Ga(1-a_i-A_is)}
{\dst\prod_{i=n+1}^p\Ga(a_i+A_is)\prod_{i=m+1}^q\Ga(1-b_i-B_is)} .
\label{Hkrn}
\end{equation}
In (\ref{Hdef}) $z^{-s}=\exp\{-s\log|z|-i\arg z\}$ and $\arg z$
is not necessarily the principal value.
The integers $m,n,p,q$ must satisfy
\begin{equation}
0\leq m \leq q, \qquad 0\leq n \leq p
\end{equation}
and empty products are interpreted as being unity.
The parameters are restricted by the condition
\begin{equation}
\bP_a\cap\bP_b=\emptyset
\end{equation}
where
\begin{align}
\bP_a &= \{\text{poles of }\Ga(1-a_i-A_is)\}
= \left\{\frac{1-a_i+k}{A_i}\in\bC: i=1,\ldots,n;k\in\bN_0\right\}
\nonumber\\
\bP_b &= \{\text{poles of }\Ga(b_i+B_is)\}
= \left\{\frac{-b_i-k}{B_i}\in\bC: i=1,\ldots,m;k\in\bN_0\right\}
\end{align}
are the poles of the numerator in (\ref{Hkrn}).
The integral converges if one of the following conditions holds
\cite{PBM90}
\begin{subequations}
\begin{align}
\Lc & = \Lc(c-i\infty,c+i\infty;\bP_a,\bP_b);\quad
|\arg z|<C\pi/2; \quad C>0 \\
\Lc &= \Lc(c-i\infty,c+i\infty;\bP_a,\bP_b);\quad
|\arg z|=C\pi/2; \quad C\geq 0; \quad cD<-\Rea F
\end{align}
\end{subequations}
\begin{subequations}
\label{minusinf}
\begin{align}
\Lc &= \Lc(-\infty+i\ga_1,-\infty+i\ga_2;\bP_a,\bP_b); 
\quad D>0; \quad 0<|z|<\infty\\
\Lc &= \Lc(-\infty+i\ga_1,-\infty+i\ga_2;\bP_a,\bP_b); 
\quad D=0; \quad 0<|z|<E^{-1}\\
\Lc &= \Lc(-\infty+i\ga_1,-\infty+i\ga_2;\bP_a,\bP_b); 
\quad D=0; \quad |z|=E^{-1}; C\geq 0; \Rea F < 0
\end{align}
\end{subequations}
\begin{subequations}
\begin{align}
\Lc &= \Lc(\infty+i\ga_1,\infty+i\ga_2;\bP_a,\bP_b); 
\quad D<0; \quad 0<|z|<\infty\\
\Lc &= \Lc(\infty+i\ga_1,\infty+i\ga_2;\bP_a,\bP_b); 
\quad D=0; \quad |z|>E^{-1}\\
\Lc &= \Lc(\infty+i\ga_1,\infty+i\ga_2;\bP_a,\bP_b);
\quad D=0; \quad |z|=E^{-1}; C\geq 0; \Rea F < 0 
\end{align}
\end{subequations}
where $\ga_1<\ga_2$.
Here $\Lc(z_1,z_2;\bG_1,\bG_2)$ denotes a contour in the
complex plane starting at $z_1$ and ending at $z_2$ and
separating the points in $\bG_1$ from those in $\bG_2$,
and the notation
\begin{align}
C & = \sum_{i=1}^nA_i - \sum_{i=n+1}^p A_i + 
\sum_{i=1}^m B_i - \sum_{i=m+1}^q B_i\\
D & = \sum_{i=1}^q B_i - \sum_{i=1}^p A_i\\
E & = \prod_{i=1}^p A^{A_i}_i\prod_{i=1}^q B^{-B_i}_i\\
F & = \sum_{i=1}^qb_i - \sum_{i=1}^p a_j + (p-q)/2 + 1
\end{align}
was employed.
The $H$-functions are analytic for $z\neq 0$ and multivalued
(single valued on the Riemann surface of $\log z$).

A change of variables in (\ref{Hdef}) shows
\begin{equation}
H^{m,n}_{p,q}\left(z
\HH{(a_1,A_1),\ldots,(a_p,A_p)}{(b_1,B_1),\ldots,(b_q,B_q)}
\right)
=
H^{n,m}_{q,p}\left(\frac{1}{z}
\HH{(1-b_1,B_1),\ldots,(1-b_q,B_q)}{(1-a_1,A_1),\ldots,(1-a_p,A_p)}
\right)
\label{Hreciprocal}
\end{equation}
which allows to transform an $H$-function with $D>0$ and $\arg z$ to one
with $D<0$ and $\arg(1/z)$.
For $\ga>0$
\begin{equation}
\frac{1}{\ga}H^{m,n}_{p,q}\left(z
\HH{(a_1,A_1),\ldots,(a_p,A_p)}{(b_1,B_1),\ldots,(b_q,B_q)}
\right) 
=
H^{m,n}_{p,q}\left(z^\ga
\HH{(a_1,\ga A_1),\ldots,(a_p,\ga A_p)}{(b_1,\ga B_1),\ldots,(b_q,\ga B_q)}
\right)
\label{Hscalarmult}
\end{equation}
while for $\ga\in\bR$
\begin{equation}
z^\ga H^{m,n}_{p,q}\left(z
\HH{(a_1,A_1),\ldots,(a_p,A_p)}{(b_1,B_1),\ldots,(b_q,B_q)}
\right) 
=
H^{m,n}_{p,q}\left(z
\HH{(a_1+\ga A_1,A_1),\ldots,(a_p+\ga A_p,A_p)}
{(b_1+\ga B_1,B_1),\ldots,(b_q+\ga B_q,B_q)}
\right)
\label{Hpowermult}
\end{equation}
holds.


\end{document}